\documentclass[conference]{IEEEtran}
\IEEEoverridecommandlockouts
\usepackage{cite}
\usepackage{amsmath}
\usepackage[qm]{qcircuit}
\usepackage{braket}
\usepackage{amsmath,amssymb,amsfonts,bm}
\usepackage{algorithmic}
\usepackage{graphicx}
\usepackage{textcomp}
\usepackage{xcolor}
\usepackage{comment}
\usepackage[bookmarks = true, pdfpagemode = None, pdfstartview = FitH, colorlinks = true, urlcolor = blue]{hyperref}
\graphicspath{{./figs/}}

\newcommand\Erase{\bgroup\markoverwith{\textcolor{red}{\rule[.5ex]{2pt}{0.4pt}}}\ULon}
\newcommand\Eraseadd{\bgroup\markoverwith{\textcolor{red}{\rule[.5ex]{2pt}{0.4pt}}}\ULon}
\newcommand{\CenterRow}[2]{
  \dimen0=\ht\strutbox%
  \advance\dimen0\dp\strutbox%
  \multiply\dimen0 by#1%
  \divide\dimen0 by2%
  \advance\dimen0 by-.5\normalbaselineskip%
  \raisebox{-\dimen0}[0pt][0pt]{#2}}

\usepackage{tikz}
\newcommand\copyrighttext{
  \footnotesize \textcopyright 2023 IEEE.  Personal use of this material is permitted.  Permission from IEEE must be obtained for all other uses, in any current or future media, including reprinting/republishing this material for advertising or promotional purposes, creating new collective works, for resale or redistribution to servers or lists, or reuse of any copyrighted component of this work in other works.}
\newcommand\copyrightnotice{
\begin{tikzpicture}[remember picture,overlay]
\node[anchor=south,yshift=10pt] at (current page.south) {\fbox{\parbox{\dimexpr\textwidth-\fboxsep-\fboxrule\relax}{\copyrighttext}}};
\end{tikzpicture}%
}

\def\BibTeX{{\rm B\kern-.05em{\sc i\kern-.025em b}\kern-.08em
    T\kern-.1667em\lower.7ex\hbox{E}\kern-.125emX}}
\begin{document}

\copyrightnotice

\title{Experimental Demonstration of Fermionic QAOA\\ with One-Dimensional Cyclic Driver Hamiltonian
}

\author{\IEEEauthorblockN{Takuya Yoshioka}
\IEEEauthorblockA{\textit{TIS Inc.} \\
2-2-1 Toyosu, Koto-ku, Tokyo 135-0061, Japan\\
email: yoshioka.takuya@tis.co.jp}
\and
\IEEEauthorblockN{Keita Sasada}
\IEEEauthorblockA{\textit{TIS Inc.} \\
2-2-1 Toyosu, Koto-ku, Tokyo 135-0061, Japan\\
}
\and
\IEEEauthorblockN{Yuichiro Nakano}
\IEEEauthorblockA{\textit{Osaka University} \\
1-2 Machikaneyama, Toyonaka, Osaka 560-0043, Japan\\
}
\and
\IEEEauthorblockN{Keisuke Fujii}
\IEEEauthorblockA{\textit{Osaka University} \\
1-2 Machikaneyama, Toyonaka, Osaka 560-0043, Japan\\
}  
\IEEEauthorblockA{\textit{RIKEN} \\
2-1 Hirosawa, Wako, Saitama 351-0198, Japan\\
}
}

\maketitle

\begin{abstract}
Quantum approximate optimization algorithm (QAOA) has attracted much attention as an algorithm that has the potential to efficiently solve combinatorial optimization problems.
Among them, a fermionic QAOA (FQAOA) for solving constrained optimization problems has been developed [Yoshioka, Sasada, Nakano, and Fujii, Phys. Rev. Research vol. 5, 023071, 2023].
In this algorithm, the constraints are essentially imposed as fermion number conservation at arbitrary approximation level.
We take the portfolio optimization problem as an application example and propose a new driver Hamiltonian
on an one-dimensional cyclic lattice.
Our FQAOA with the new driver Hamiltonian reduce the number of gate operations in quantum circuits.
Experiments on a trapped-ion quantum computer using 16 qubits on Amazon Braket
demonstrates that the proposed driver Hamiltonian effectively suppresses noise effects compared to the previous FQAOA.

\end{abstract}


\section{Introduction}\label{sec:intro}
Quantum optimization algorithms are attracting much attention as a solution to combinatorial optimization problems in various industries \cite{Rosenberg, Hodson, Ajagekar, Streif1, Harwood, Gilliam, Niroula, Nikmehr}.
Quantum annealing based on adiabatic theorems is the standard quantum approach used to solve the optimization problems \cite{Kadowaki, Farhi1},
however, it is limited to quadratic, unconstrained binary optimization (QUBO) problems.
In particular, it is known that converting constrained combinatorial optimization problems to the QUBO form can lead to problems in computational accuracy.
On the other hand, quantum approximate optimization algorithm (QAOA) \cite{Farhi2},
has been proposed as a hybrid quantum classical algorithm for solving complex combinatorial optimization problems using a universal quantum computer.
Subsequently, quantum alternating operator ansatz \cite{Hadfield1, Wang} and fermionic QAOA (FQAOA) \cite{FQAOA} have been developed for constrained optimization problems.
In these frameworks, the hard constraints are imposed, so that the constraints are strictly satisfied.
In particular, the FQAOA has the characteristic of guaranteed convergence because it is reduced to a quantum adiabatic algorithm (QAA) in the limit of large circuit depth $p$.
The application of FQAOA to the portfolio optimization problem confirms its clear advantage over previous study in Ref. \cite{Hodson}.

In the previous FQAOA study \cite{FQAOA},
the combinatorial optimization problem under Hamming weight constant constraint has been recast as an energy minimization problem under the constant number of fermions.
The FQAOA ansatz satisfying the hard constraints is written using the variational parameters (${\bm \gamma},{\bm \beta}$) as follows:
\begin{equation}
  \ket{\psi_p({\bm \gamma},{\bm \beta})}=\left[\prod_{j=1}^{p}\hat{U}_m(\beta_j)\hat{U}_p(\gamma_j)\right]\hat{U}_{\rm init}\ket{\rm vac},  \label{eq:Ansqaoa}
\end{equation}
where $\hat{U}_p(\gamma) = e^{-i\gamma\hat{\cal H}_{p}}$ is a phase rotation unitary following problem Hamiltonian $\hat{\cal H}_p$,
$\hat{U}_m(\beta) = e^{-i\beta\hat{\cal H}_d}$ is a mixing unitary following driver Hamiltonian $\hat{\cal H}_d$,
and $\hat{U}_{\rm init}$ is a unitary preparing an initial state $\ket{\phi_0}$ as  $\ket{\phi_0}=\hat{U}_{\rm init}\ket{\rm vac}$, where $\ket{\rm vac}$ is a vacuum without fermions.
The hard constraint condition can be written as follows:
\begin{equation}
  \hat{C}\ket{\psi_p({\bm \gamma},{\bm \beta})}=M\ket{\psi_p({\bm \gamma},{\bm \beta})},\label{eq:const}
\end{equation}
where $\hat{C}$ is a total particle number operator.
The driver Hamiltonian $\hat{\cal H}_d$ and the initial state $\ket{\phi_0}$ are assumed to satisfy the design guidelines in Ref. \cite{FQAOA}.
This is restated here
\begin{itemize}
\item Condition I
\begin{itemize}    
 \item$[\hat{\cal H}_d, \hat{C}]=0$.
\end{itemize}  
\item Condition II
  \begin{itemize}
 \item $|\bra{\phi_{{\bm x}'}}(\hat{\cal H}_d)^n\ket{\phi_{\bm x}}| >0,$
\end{itemize}
  for a sufficiently large value of $n$, where $\ket{\phi_{\bm x}}$ and $\ket{\phi_{\bm x'}}$ are arbitrary eigenstates of $\hat{\cal H}_p$
  that satisfy the constraints.
\item Condition III
\begin{itemize}
  \item $\hat{\cal H}_d\ket{\phi_0} = E_{0}\ket{\phi_0},\label{eq:Hpeigen}$
  \item $\hat{C}\ket{\phi_0}=M\ket{\phi_0},\label{eq:qconstphi0}$
\end{itemize}
where $E_0$ is the ground state energy of $\hat{\cal H}_d$.
\end{itemize}
The conditions I and II restrict the candidates for $\hat{\cal H}_d$.
Although the previous study have employed tight-binding models on $D$-leg ladder lattice \cite{FQAOA},
$\hat{\cal H}_d$ satisfying the condition is not limited to this.
Therefore, it is important to design a driver Hamiltonian that
minimizes the influence of noise when using an actual quantum processing unit (QPU).

In this study, we propose a new driver Hamiltonian on an one-dimensional cyclic lattice in the FQAOA framework
and show that its implementation in a quantum computer can reduce the influence of noise.
First, to clarify the impact of the new driver Hamiltonian,
we perform noiseless simulations of the FQAOA on cyclic lattice and
show that its performance is comparable to that of the previous FQAOA on the ladder lattice \cite{FQAOA}.
Experiments using a trapped-ion quantum computer demonstrate that the proposed FQAOA outperforms the previous one by reducing errors caused by gate operations in quantum circuits.

This paper is organized as follows.
In section \ref{sec:Hdphi}, we propose a new cyclic driver Hamiltonian with low computational cost and explicitly indicate its ground state.
Section \ref{sec:port} takes the formulation of the portfolio optimization problem as an example of a constrained optimization problem and shows how the new ansatz can be implemented on quantum circuits.
Section \ref{sec:res} presents the results of simulations and experiments with FQAOA using the new driver Hamiltonian.
Section \ref{sec:sum} provides a summary and discussion.

\section{One-Dimensional Cyclic Driver Hamiltonian and Ground State} \label{sec:Hdphi}

\begin{figure}[htb]
  \includegraphics[width=8.5cm]{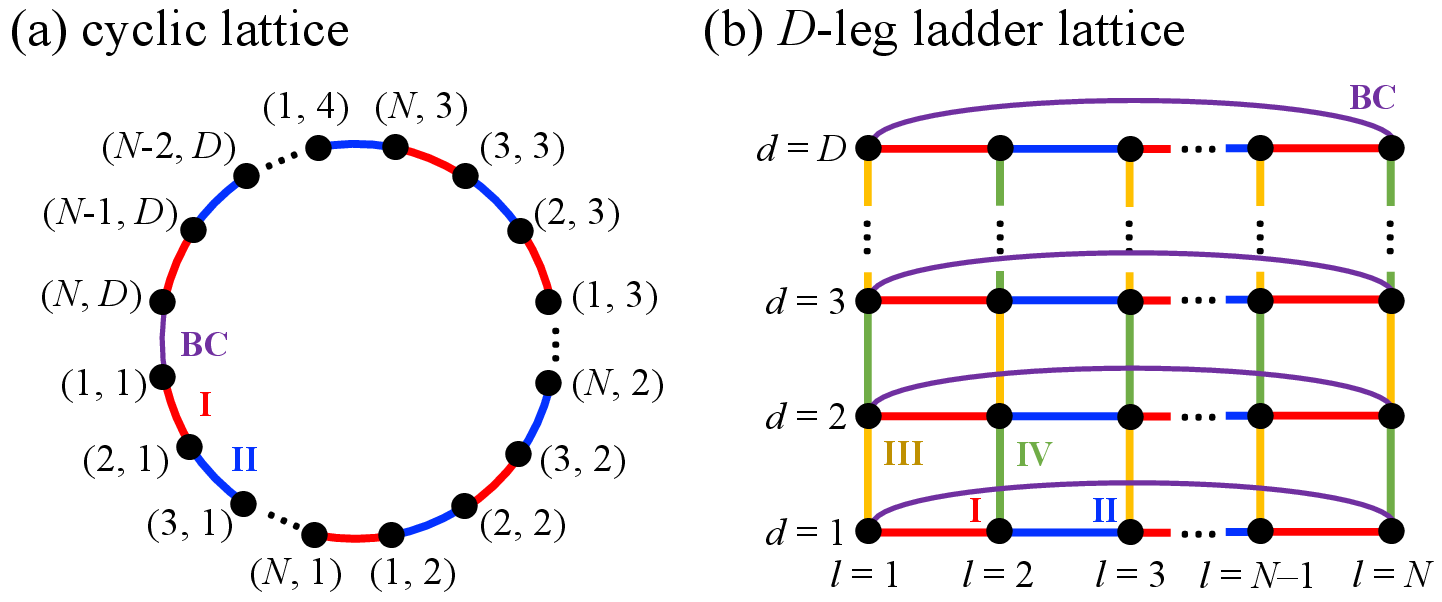}
  \caption{\label{fig:lattice}
    Lattice structure of driver Hamiltonian $\hat{\cal H}_{\rm hop}^{\alpha}$ for $\alpha=$ (a) cyc and (b) lad.
    The hopping terms of $\hat{U}_{\delta}^{\rm lad (cyc)}(\beta)$ ($\delta$ = I, II, III, IV, and BC),
    which decomposes $\hat{U}_m$, is indicated in different colors (see text). 
  }
\end{figure}

We propose a new driver Hamiltonian $\hat{\cal H}_d=\hat{\cal H}_{\rm hop}^{\rm cyc}$ with low computational cost and its ground states $\ket{\phi_0} = \ket{\phi_0^{\rm cyc}}$
, which is initial state of FQAOA, 
  following the guideline shown in section \ref{sec:intro} and in Ref \cite{FQAOA}.
To this end, we first introduce the fermionic formulation in unary encoding and constraint condition, and then
we present the tight-binding model on an one-dimensional cyclic lattice shown in Fig. \ref{fig:lattice} (a) as the new driver Hamiltonian.
We also show the driver Hamiltonian $\hat{\cal H}_d=\hat{\cal H}_{\rm hop}^{\rm lad}$ on the $D$-leg ladder lattice [Fig. \ref{fig:lattice} (b)] used in the previous study \cite{FQAOA}
and its ground state $\ket{\phi_0^{\rm lad}}$ in APPENDIX \ref{sec:aplad}.
Note that the following arguments do not depend on the details of a problem Hamiltonian $\hat{\cal H}_p$ with unary encoding.

\subsection{Mapping to Fermionic Formulation}
In unary encoding, any integer $z_l\in\{0, 1, \cdots D\}$ for $l=0, 1, \cdots N$ can be written by using binary $x_{l,d}\in\{0, 1\}$ as $z_l= \sum_{d=1}^Dx_{l,d}$.
In the fermionic formulation, the binary variable is mapped to the number operator of fermions as $x_{l,d}\mapsto\hat{n}_{l,d}$ and
\begin{equation}
  \hat{n}_{l,d}\ket{\phi_{\bm x}} =x_{l, d}\ket{\phi_{\bm x}},
\end{equation}
with the computational basis
\begin{equation}
  \ket{\phi_{\bm x}}=\prod_{l=1}^{N}\prod_{d=1}^D\left(\hat{c}^{\dagger}_{l,d}\right)^{x_{l,d}}\ket{\rm vac}, \label{eq:base}
\end{equation}
where $\hat{n}_{l,d}=\hat{c}^\dagger_{l,d}\hat{c}_{l,d}$ and $\hat{c}^\dagger_{l,d}$($\hat{c}_{l,d}$) is the creation (annihilation) operator of fermion.

\subsection{Constraint on Constant Hamming Weight}
In the following, we consider the constraint that the sum of integer variables $z_l$ is a constant value $M$ as:
\begin{equation}
 \sum_{l=1}^N z_l = \sum_{l=1}^N\sum_{d=1}^D x_{l,d}=M.
\end{equation}
Using the operator $\hat{C}$ of the constraint in the fermionic formulation,
the constraint condition can be written as the following eigenvalue equation:
\begin{equation}
  \hat{C}\ket{\phi_{\bm x}}= M\ket{\phi_{\bm x}}, \label{eq:CM}
\end{equation}
with
\begin{equation}
 \hat{C}= \sum_{l=1}^N\sum_{d=1}^D \hat{n}_{l,d}.
\end{equation}  
Therefore, the constraint in Eq. (\ref{eq:CM}) corresponds to the conservation law of the number of fermions.

\subsection{Cyclic Driver Hamiltonian and Ground States}
Here we construct a simple tight-binding model as the low cost driver Hamiltonian $\hat{\cal H}_d=\hat{\cal H}_{\rm hop}^{\rm cyc}$
according to the following directions.
First, from conditions I,  we assume that $\hat{\cal H}_d$ is a tight binding model consisting of hopping terms of fermions.
Next, from condition II, we design $\hat{\cal H}_d$ such that all sites labeled with ($l,d$) are connected by the hopping terms.
Here, we develop a lattice geometry such that the number of hopping terms is minimized. 
Finally, from condition III, we express the ground state of $\hat{\cal H}_d$ as a single Slater determinant satisfying the constraint.


The tight-binding model on a cyclic lattice shown in Fig. \ref{fig:lattice} (a) is defined as:
\begin{align}
  \hat{\cal H}_{\rm hop}^{\rm cyc}=&-t\sum_{l=1}^{N-1}\sum_{d=1}^{D}\left(\hat{c}_{l,d}^{\dagger}\hat{c}_{l+1,d}+\hat{c}_{l+1,d}^{\dagger}\hat{c}_{l,d}\right)\nonumber\\
  &-t\sum_{d=1}^{D-1}\left(\hat{c}_{1,d+1}^{\dagger}\hat{c}_{N,d}+\hat{c}_{N,d}^{\dagger}\hat{c}_{1,d+1}\right)\nonumber\\
  &-t(-1)^{M-1}\left(\hat{c}_{1,1}^{\dagger}\hat{c}_{N,D}+\hat{c}_{N,D}^{\dagger}\hat{c}_{1,1}\right)\nonumber\\  
  =&\sum_{k=1}^{ND}\varepsilon^{\rm cyc}_{k+\delta}\hat{\alpha}^{\dagger}_{k+\delta}\hat{\alpha}_{k+\delta},
  \label{eq:Hcyc}
\end{align}\\
where $t$ is the hopping integral in the tight-binding model,
which has a periodic (anti-periodic) boundary condition due to the odd (even) number of $M$.
The lattice labels $(l,d)$ are shown in Fig. \ref{fig:lattice} (a).
The $\varepsilon^{\rm cyc}_{k+\delta}$ in the last line of Eq. (\ref{eq:Hcyc}) is single-particle energy,
where $\delta=-0.5$ $(0)$ for an even (odd) number of fermions $M$.
This $\hat{\cal H}_{\rm hop}^{\rm cyc}$ corresponds to a one-dimensional $XY$-Hamiltonian by the Jordan-Wigner transformation \cite{Jordan, Lieb}.

The energy dispersion $\varepsilon_{q}^{\rm cyc}$ and the operator $\hat{\alpha}_{q}$ in Eq. (\ref{eq:Hcyc}) have the following form:
\begin{equation}
  \varepsilon^{\rm cyc}_{q}=-2t\cos\left(\frac{2\pi q}{ND}\right), \label{eq:ekcyc}
\end{equation}
and
\begin{equation}  
  \hat{\alpha}_{q}^\dagger = \sum_{l=1}^{N}\sum_{d=1}^{D}[{\bm \phi}_0^{\rm cyc}]_{q,(l,d)}\hat{c}_{l,d}^{\dagger},
\end{equation}
respectively, with
\begin{equation}
\ [{\bm \phi}_0^{\rm cyc}]_{q,(l,d)}\\
=  \left \{\begin{alignedat}{2}
  a_q\sin &\left(
  \frac{2\pi q}{ND}[l+N(d-1)]\right), \\
  &\text{if } 0< q < ND/2,\\ 
  a_q\cos &\left(
  \frac{2\pi q}{ND}[l+N(d-1)]\right),\\
  &\text{if } ND/2\le q \le ND,\\
\end{alignedat}
\right.
  \label{eq:phi0cyc}  
\end{equation}
where $q = k+\delta$, and normalization factors $a_q = \sqrt{1/ND}$ for $q=ND/2$ and $ND$,
and $a_q = \sqrt{2/ND}$ for the other $q$'s.
Finally, The ground state of $\hat{\cal H}_d=\hat{\cal H}_{\rm hop}^{\rm cyc}$,
which is the initial state of FQAOA,
is explicitly written as follows:
\begin{equation}
\ket{\phi_0^{\rm cyc}}\\
=  \left \{\begin{alignedat}{2}
&\prod_{k=1}^{M/2}\hat{\alpha}_{k-1/2}^{\dagger}
\hat{\alpha}_{ND-k+1/2}^{\dagger}\ket{\rm vac}
  &\ &\text{if $M$ is even},\\ 
&\hat{\alpha}_{ND}^{\dagger}\prod_{k=1}^{(M-1)/2}\hat{\alpha}_{k}^{\dagger}\hat{\alpha}_{ND-k}^{\dagger}
\ket{\rm vac}
  &\ &\text{if $M$ is odd},
\end{alignedat}
\right.
  \label{eq:phi0cycket}
\end{equation}
which is uniquely determined without degeneracy.

\section{Portfolio Optimization Problem}\label{sec:port}
As an application of FQAOA using the proposed driver Hamiltonian $\hat{\cal H}_d=\hat{\cal H}_{\rm hop}^{\rm cyc}$, 
we take portfolio optimization problems \cite{FQAOA, Markowitz, Hodson, Rosenberg} and evaluate its performance by comparing with previous FQAOA study
using $\hat{\cal H}_d=\hat{\cal H}_{\rm hop}^{\rm lad}$ \cite{FQAOA} and $XY$-QAOA study using $XY$ driver Hamiltonian \cite{Hodson}.

\subsection{Problem Hamiltonian}
In this study for the portfolio optimization problem,
the cost function $E({\bm x})$ for given bit string ${\bm x} \in \{0, 1\}^{ND}$
and the constraint condition of the total number of stock holdings being $K$
are those defined by Eq. (30) and (31) of Ref. \cite{FQAOA}, respectively.
The minimum eigenvalue problem for the portfolio optimization problem can be written as \cite{FQAOA}:
\begin{align}
  \hat{\cal H}_p\ket{\phi_{\bm x}}&=E({\bm x})\ket{\phi_{\bm x}},\\
  \hat{C}\ket{\phi_{\bm x}}&=M\ket{\phi_{\bm x}},
  \label{eq:Hconst}
\end{align}
where $\ket{\phi_{\bm x}}$ is in Eq. (\ref{eq:base}) and $M=ND/2-K$.
The operators $\hat{\cal H}_p$ and $\hat{C}$ are written as follows:
\begin{align}
  \hat{\cal H}_p=& \frac{\lambda}{K^2}\sum_{l,l'=1}^N \sigma_{l,l'}
  \sum_{d,d'=1}^D\left(\hat{n}_{l, d}-\frac{1}{2}\right)\left(\hat{n}_{l', d'}-\frac{1}{2}\right)\nonumber\\
  &+\frac{1-\lambda}{K}\sum_{l=1}^N\mu_l\sum_{d=1}^D\left(\hat{n}_{l,d}-\frac{1}{2}\right),
  \label{eq:Hpport}\\
  \hat{C}=&\sum_{l=1}^N\sum_{d=1}^D\hat{n}_{l,d},
  \label{eq:Nport}
\end{align}
where $\sigma_{l,l'}$, $\mu_l$ and $\lambda$ denote the asset covariance,
the average return, and the risk capacity parameter of the asset manager, respectively. 

The objective is to obtain a bit string that gives the lowest cost under the constraints; in other words,
in the framework of FQAOA, a ground state of the fermion system of $M$-particles.
For this purpose, expectation values of energy $E_p({\bm \gamma},{\bm \beta})$
are calculated, which is obtained as the output of the FQAOA ansatz in Eq. (\ref{eq:Ansqaoa}) as follows:
\begin{equation}
  E_p({\bm \gamma},{\bm \beta})=\braket{\psi_p({\bm \gamma},{\bm \beta})|\hat{\cal H}_p|\psi({\bm \gamma},{\bm \beta})}. \label{eq:Ep}
\end{equation}
The parameters $({\bm \gamma},{\bm \beta})$ can be estimated from the time-discretized QAA \cite{FQAOA} as:
\begin{equation}
  \gamma_j^{(0)} = \frac{2j-1}{2p}\Delta t,\ \ \ \beta_j^{(0)}  = \left(1-\frac{2j-1}{2p}\right)\Delta t,\label{eq:gamma0beta0}
\end{equation}
where $\Delta t$ is a unit of discretized time.
The derivation of which is described in APPENDIX \ref{sec:aplad} in Ref. \cite{FQAOA}.
As with other variational algorithms, these parameters are optimized according to the following equation:
\begin{equation}
  E_p({\bm \gamma^*},{\bm \beta}^*)=\min_{{\bm \gamma},{\bm \beta}}E_p({\bm \gamma},{\bm \beta}), \label{eq:Ep_opt}
\end{equation}
where ${\bm \gamma^*}$ and ${\bm \beta}^*$ give the minimum expectation value of energy.

\subsection{Implementation on Quantum Circuit}
In this subsection,
we describe how to implement FQAOA ansatz with the cyclic driver Hamiltonian $\hat{\cal H}_{\rm hop}^{\rm cyc}$ on quantum circuits.
The phase rotation unitary $\hat{U}_p(\gamma)$
and initial states preparation unitary $\hat{U}^{\rm cyc}_{\rm init}$ can be implemented as in the case on the ladder lattice \cite{FQAOA}.
Therefore, we specifically describe an implementation of mixing unitary $\hat{U}^{\rm cyc}_m(\beta)=\exp({-i\beta\hat{\cal H}_{\rm hop}^{\rm cyc}})$.

The mixing unitary $\hat{U}_m^{\rm cyc}(\beta)$ can be explicitly written as:
\begin{equation}
  \hat{U}^{\rm cyc}_m(\beta)=\hat{U}^{\rm cyc}_{\rm BC}(\beta)\hat{U}^{\rm cyc}_{\rm II}(\beta)\hat{U}^{\rm cyc}_{\rm I}(\beta),
\end{equation}
with
\begin{align}
  \hat{U}^{\rm cyc}_{\rm I(II)}(\beta)&=\prod_{i\ {\rm odd(even)}}\exp\left[i\beta t\left(\hat{c}^{\dagger}_i\hat{c}_{i+1}+\hat{c}^{\dagger}_{i+1}\hat{c}_{i}\right)\right],\label{eq:ophop}\\  
  \hat{U}^{\rm cyc}_{\rm BC}(\beta)&=\exp\left[i\beta t(-1)^{M-1}\left(\hat{c}^{\dagger}_{ND}\hat{c}_{1}+\hat{c}^{\dagger}_{1}\hat{c}_{ND}\right)\right],
  \label{eq:ophopbc}
\end{align}
where $M=ND/2-K$ and
the subscript of $\hat{c}^{(\dagger)}_{i}$ is the sequential number $i = l+N(d-1)$ along the circle shown in Fig. \ref{fig:lattice} (a).

The quantum circuit corresponding to each hopping pair in Eqs. (\ref{eq:ophop}) and (\ref{eq:ophopbc}) is as follows:
\begin{align}
  \begin{split}
    \Qcircuit @C=0.5em @R=.7em {
      &&\lstick{a}& \gate{R_x({-\pi/2})} &\ctrl{1}& \gate{R_x(-\beta t)} &\ctrl{1} & \gate{R_x({\pi/2})}  & \qw \\
      &&\lstick{b}& \gate{R_x({\pi/2})}  &\targ   & \gate{R_z(\beta t)}  &
\targ    & \gate{R_x({-\pi/2})} &\qw&,
}
\end{split}\label{eq:Qchop}
\end{align}
where $(a,b)=(i, i+1)$ for $\hat{U}^{\rm cyc}_{\rm I(II)}$ and $(a, b)=(1, ND)$ for $\hat{U}^{\rm cyc}_{\rm BC}$.
This circuit is equivalent to that of the $XY$ gate.

As a result, the number of quantum gate operations in $\hat{U}^{\rm cyc}_m(\beta)$ is reduced from $O(N^2D)$ to $O(ND)$ compared to that
in the preceding study \cite{FQAOA}.
This is because all hopping terms on the cyclic lattice follow the Jordan-Wigner ordering,
thus avoiding the fermionic swap gates shown in Ref. \cite{FQAOA}.
The number of gates required for FQAOA ansatz is summarized in TABLE \ref{tble:gate}.

\begin{table}[htb]
  \caption{\label{tble:gate}
    Numbers of single- and two-qubit gates in $\hat{U}^{\alpha}_{\rm init}$, $\hat{U}_p$,
    and $\hat{U}^{\alpha}_m$ of $\alpha$-FQAOA ansatz for $\alpha$ = cyc and lad.
    Here, the two qubit gates represent the CNOT and the SWAP gates.
    The Givens rotation gate in $\hat{U}^{\rm cyc, lad}_{\rm init}$ used in this study is shown in APPENDIX \ref{sec:apgivens}.
  }
  \begin{center}
{\renewcommand\arraystretch{2.0}
  \begin{tabular}{|c|c|c|}
    \hline
    operator &single-qubit gate& two-qubit gate\\
    \hline
    $\hat{U}_{\rm init}^{\rm cyc, lad}$&
$\displaystyle{\frac{1}{2}(4ND+8K+1)(ND-2K)}$     
    &
    $\displaystyle{\frac{1}{2}\left[(ND)^2-4K^2\right]}$\\
        \hline
$\hat{U}_{p}$  & $\displaystyle{\frac{1}{2}ND(ND+1)}$
        & $ND(ND-1)$    \\
            \hline
            $\hat{U}_{m}^{\rm cyc}$ & $6ND$ & $2ND$\\
            $\hat{U}_{m}^{\rm lad}$ & $2N^2D+10ND-6N$ & $2N^2D+2ND-2N$\\            
            \hline
\end{tabular}}
\end{center}
\end{table}

\section{Results} \label{sec:res}
This paper takes up the portfolio optimization problems and compares and
evaluates the performance of FQAOA using the newly proposed cyclic Hamiltonian $\hat{\cal H}_{\rm hop}^{\rm cyc}$
with that using the ladder Hamiltonian $\hat{\cal H}_{\rm hop}^{\rm lad}$ from both in terms of noiseless simulations and QPU experiments.
Hereafter, the FQAOA calculation using $\hat{\cal H}_d=\hat{\cal H}_{\rm hop}^{\alpha}$ for ($\alpha$ = cyc and lad) will be referred to as $\alpha$-FQAOA.

\subsection{Computational Details}

Numerical simulations have been performed using the fast quantum simulator qulacs \cite{qulacs}.
The parameters of the fixed-angle FQAOA are those determined by the QAA shown in Eq. (\ref{eq:gamma0beta0}).
The optimized parameters and energies in Eq. (\ref{eq:Ep_opt})
is obtained by the Broyden-Fletcher-Goldfarb-Shanno (BFGS) method using $({\bm \gamma}^{(0)}, {\bm \beta}^{(0)})$ as the initial angle.

The actual quantum computations are performed using
the IonQ device Aria \cite{IonQ}, provided by Amazon Braket \cite{Amazon}.
The variational parameters of FQAOA used in the QPU experiments are adapted from those determined by noiseless simulations.
For each probability distribution, the error-mitigation technique, debiasing \cite{Amazon_em, error-mitigation} is applied to 10,000 measurements.
Note that the sharpening technique \cite{Amazon_em} is not applied in this paper.

\subsection{Model Parameters}

For model parameters of $\hat{\cal H}_p$ in Eq. (\ref{eq:Hpport}), $\sigma_{l,l'}$ and $\mu_l$ are taken from Fig. 2 and TABLE IV of the paper \cite{Hodson}, respectively,
and $\lambda=0.9$, $N=8$, $D=2$, and $K=4$.
These parameters are the same as in previous studies \cite{Hodson} and \cite{FQAOA}.
$W$ is the range of total energy that the $\hat{\cal H}_p$ can take under the constraint condition.
Since $W$ is independent of the type of $\hat{\cal H}_d$, it is adopted as the energy scale for performance comparison.

For model parameters of $\hat{\cal H}_d$ in Eqs. (\ref{eq:Hcyc}) and (\ref{eq:Hlad}),
the hopping integral $t$ and $t=t^{\parallel}=t^{\perp}$ are
determined from the ratio of the energy scales of $\hat{\cal H}_p$ and $\hat{\cal H}_d=\hat{\cal H}_{\rm hop}^{\alpha}$ for $\alpha$ = cyc and lad as $t=W/W_{\rm hop}^{\alpha}$,
where $W_{\rm hop}^{\alpha}$ is the range of total energy of $\hat{\cal H}_{\rm hop}^{\rm \alpha}$ at $t=1$.

\subsection{Simulation Results without Noise}

\begin{figure}[htb]
  \includegraphics[width=8.5cm]{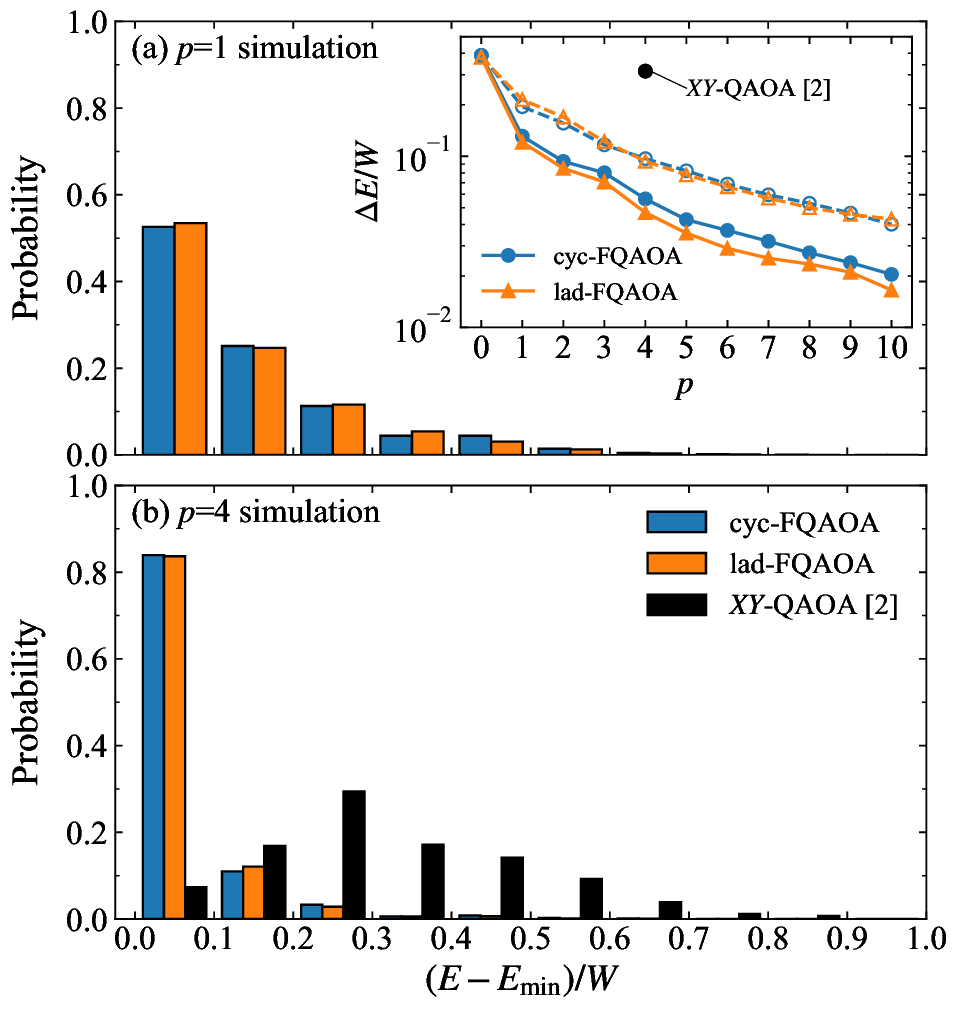}
  \caption{\label{fig:hist}
    Probability distributions of energy obtained by the noiseless simulations
    by using $\alpha$-FQAOA for $\alpha$ = cyc and lad at (a) $p$=1 and (b) $p$=4,
    where the previous $XY$-QAOA results at $p=4$ \cite{Hodson} are shown in (b).
    The horizontal axis takes the energy value measured from the ground state energy $E_{\rm min}$.
    The inset in (a) shows expectation values $\Delta E/W = [E_p({\bm \gamma}, {\bm \beta})-E_{\rm min}]/W$
    in Eq. (\ref{eq:Ep}) in log scale,
    where the results using
    $({\bm \gamma}^{(0)}, {\bm \beta}^{(0)})$ at $W\Delta t=10$ in Eq. (\ref{eq:gamma0beta0})
    and optimized $({\bm \gamma}, {\bm \beta})=({\bm \gamma}^*, {\bm \beta}^*)$ are shown
    by the open and closed plots, respectively.
  }
\end{figure}

Fig. \ref{fig:hist} shows the results of the probability distribution for energy obtained by the
$\alpha$-FQAOA for $\alpha$ = cyc and lad at (a) $p=1$ and (b) $p=4$.
The peaks appear in the same lowest energy region $0<(E-E_{\rm min})<W/10$,
and the probabilities of finding the low-energy states are almost the same for both models at $p=1$ and $p=4$.
We can also confirm that both FQAOAs at $p = 1$ outperform $XY$-QAOA at $p=4$.

From another perspective, the expectation values of energy $E_p({\bm \gamma}, {\bm \beta})$ in Eq.  (\ref{eq:Ep}) are shown in the inset of Fig. \ref{fig:hist} (a).
The fixed-angle FQAOA energies shown in the open plots are reduced to the closed plots by parameter optimization.
Focusing on the parameter-optimized FQAOA, 
the lower energies of lad-FQAOA shown by triangles are attributed to the fact that
the mixing works effectively on the ladder lattice because of the large number of hopping paths as shown in Fig \ref{fig:lattice} (b).
However, since there is a trade-off between the number of gate operations and computational performance.
In actual QPU experiments, lad-FQAOA may be more affected by noise due to the large number of gate operations.


\subsection{Experimental Results}

\begin{figure}[htb]
  \includegraphics[width=8.5cm]{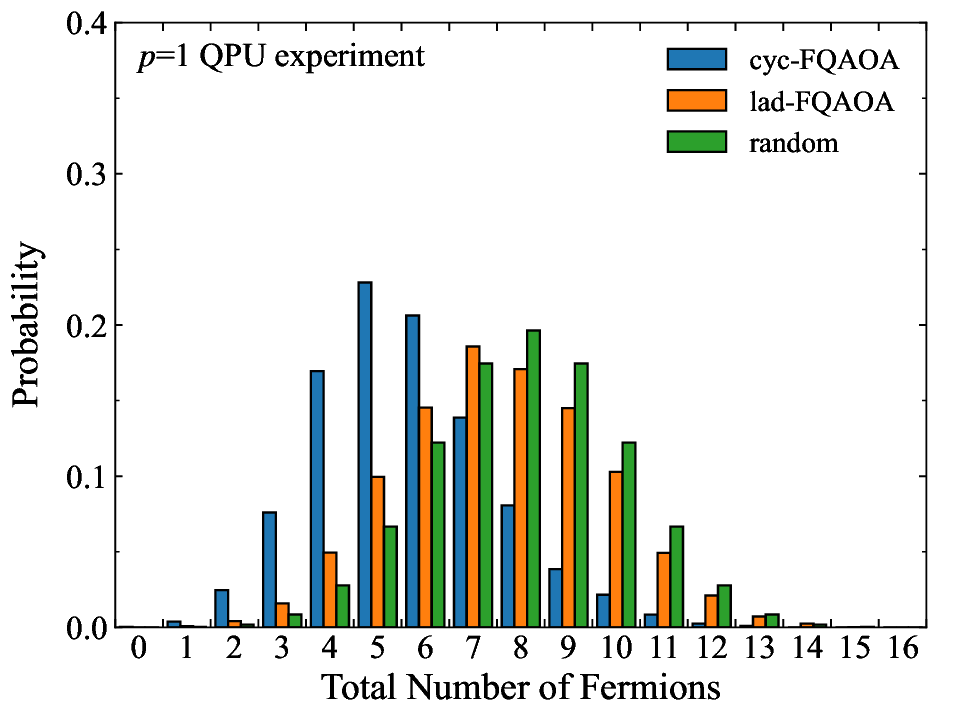}
  \caption{\label{fig:histnum}
    Probability distributions $P_M$ of total number of fermions $M$ obtained by the QPU experiments
    by using $\alpha$-FQAOA for $\alpha$ = cyc and lad at $p$=1, and random sampling represented by
    $P_M={}_{ND}C_{M}/2^{ND}$.
  }
\end{figure}

In an ideal noiseless environment, the conservation law for the fermion number $M$ expressed
in Eq. (\ref{eq:Hconst}) is strictly satisfied.
However, in experiments with actual QPUs, this conservation law may be violated.
Therefore, we first present
the results of an investigation of the probability distributions $P_M$ of the number of fermions $M$,
which expressed by the following equation:
\begin{equation}
P_M = \sum_{\bm x}|\braket{\phi_{\bm x}|\psi_p({\bm \gamma^*}, {\bm \beta^*})}|^2\delta_{M_{\bm x}, M},
\end{equation}
where $M_{\bm x}$ is the eigenvalue of $\hat{C}\ket{\phi_{\bm x}}=M_{\bm x}\ket{\phi_{\bm x}}$
which is equivalent to the Hamming weight of bit string ${\bm x}$.
Here, the constraint on the number of fermions $M = ND/2-K = 4$ is imposed,
so $P_M = \delta_{M, 4}$ in the noiseless environment.
The probability distribution $P_M$ by QPU experiment with debiasing is shown in Fig. \ref{fig:histnum}.
Compared to the lad-FQAOA, the cyc-FQAOA takes a relatively large value of $P_{M=4}=0.17$.
This difference is due to the fact that the cyclic lattice reduces the number of gate operations,
thereby reducing the effect of noise.
In the following, we show the results of post-selection, in which only $M=4$ is extracted from the QPU measurements.

\begin{figure}[htb]
  \includegraphics[width=8.5cm]{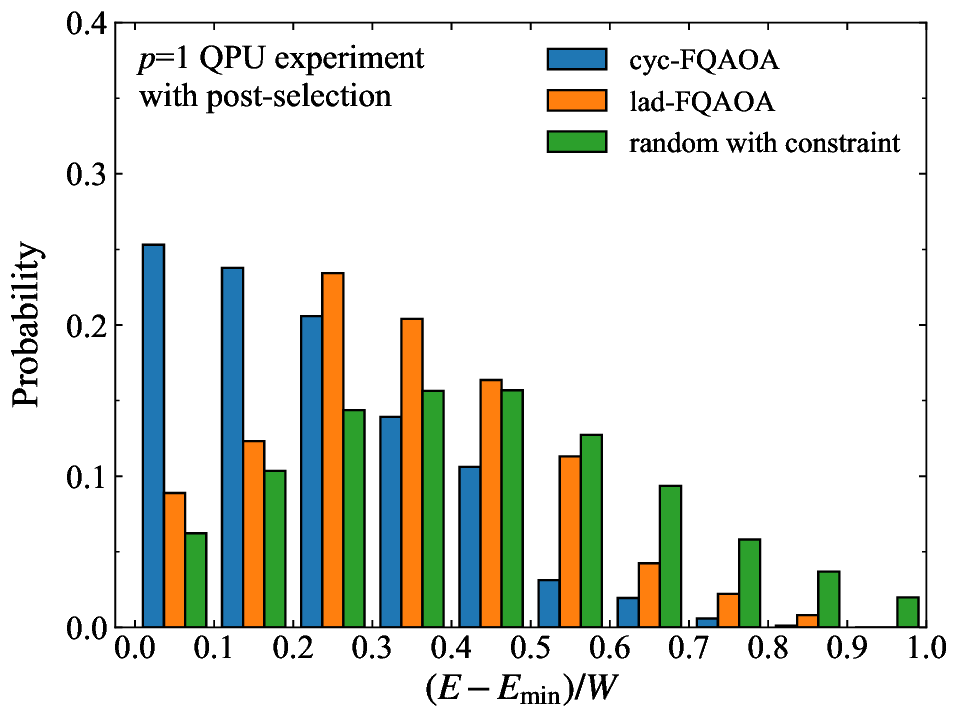}
  \caption{\label{fig:histexp}
    Probability distributions of energy obtained by the QPU experiments
    by using $\alpha$-FQAOA for $\alpha$ = cyc and lad at $p$=1 and random sampling,
    where post-selection is carried out to the samples that satisfy the constraint.
    The energy value measured from the ground state energy $E_{\rm min}$.    
  }
\end{figure}

Fig. \ref{fig:histexp} shows the experimental results of the probability distribution for energy at $p=1$
obtained by the $\alpha$-FQAOA for $\alpha$ = cyc and lad,
where the post-selection to samples satisfying the number of particles $M=4$ is applied after adopting the debiasing.
The peak appears in the lowest energy region for cyc-FQAOA,
while for lad-FQAOA it appears in $2W/10<(E-E_{\rm min})<3W/10$.
Comparing the peak values in $0<(E-E_{\rm min})<W/10$,
the value for lad-FQAOA is reduced to about 0.1 due to significant influence of noise,
while the value for cyc-FQAOA is kept at about 0.25.
This results indicate that cyc-FQAOA succeeds in reducing the influence of noise compared to lad-FQAOA.


\begin{table}[htb]
  \caption{\label{tble:exp}
    Expectation values of energy $\Delta E/W = [E_p({\bm \gamma}^*, {\bm \beta}^*)-E_{\rm min}]/W$ in Eq. (\ref{eq:Ep}) and its standard deviation $\sigma$
    by using $\alpha$-FQAOA for $\alpha$ = cyc and lad at $p=1$.
    For QPU experiments, after adopting the debiasing, post-selection is carried out to the samples that satisfy the constraint.
  }
  \begin{center}
{\renewcommand\arraystretch{2.0}
  \begin{tabular}{|c|c|c|}
    \hline
    & \multicolumn{2}{|c|}{$\Delta E/W$ ($\sigma$)}\\ \cline{2-3} 
    method           &noiseless simulation& QPU experiment\\
    \hline
    cyc-FQAOA (${p=1}$)& 0.13 (0.13) & 0.23 (0.16)$^{\mathrm{a}}$\\\hline
    lad-FQAOA (${p=1}$)& 0.12 (0.13) & 0.34 (0.17)$^{\mathrm{b}}$\\ \hline
    random with constraint& 0.38 (0.18)& ---\\ \hline
    \multicolumn{3}{l}{\raisebox{0mm}{$^{\mathrm{a}}$ Without the debiasing, the same as 0.23 (0.16).}}\\
    \multicolumn{3}{l}{\raisebox{3mm}{$^{\mathrm{b}}$ Without the debiasing, 0.33 (0.18).}}
\end{tabular}}
  \end{center}
\end{table}

The results of the energy expectation values with standard deviations are shown in TABLE \ref{tble:exp}.
The values of the lad-FQAOA and cyc-FQAOA experiments are both smaller than that of random guesses under the constraint condition.
Among them, consistent with the results in Fig. \ref{fig:histexp},
the value of cyc-FQAOA obtained by the QPU experiment is smaller than that of lad-FQAOA.
This is because in the lad-FQAOA, the negative influence of noise due to many gate operations is larger 
than the slight decrease in energy in the noise-free environment.
Indeed, the number of single- and two-qubit gate operations (Q1, Q2) required
to implement $\alpha$-FQAOA at QAOA level $p$ is larger
(Q1, Q2) = (388+504$p$, 96+512$p$) for $\alpha$ = lad
than
(Q1, Q2) = (388+232$p$, 96+272$p$) for $\alpha$ = cyc.
Notably, the energy value of 0.23 for the cyc-FQAOA experiment at $p=1$ is lower than
the value of 0.3 for the noiseless $XY$-QAOA simulation at $p=4$.
These results are robust regardless of the debiasing technique \cite{Amazon_em} applied in this study,
as can be seen from the footnote of TABLE \ref{tble:exp}.




\section{Summary and Discussion} \label{sec:sum}
We proposed a new driver Hamiltonian for the fermionic quantum approximate optimization algorithm (FQAOA)
to efficiently solve combinatorial optimization problems with constraints,
taking the portfolio optimization problem.
We showed that the new driver Hamiltonian on the one-dimensional cyclic lattice reduces the number of gate operations in quantum circuits.
Noiseless simulations with the new driver Hamiltonian revealed
that it performs as well as the previous FQAOA in both fixed-angle FQAOA and parameter-optimized FQAOA calculations.
Experiments with trapped-ion quantum processing unit (QPU) at $p=1$ demonstrated that
FQAOA on the cyclic lattice improves the probability of obtaining low-cost solutions by a factor of 2.5 compared to the ladder lattice.

Post-selection applied in this study worked well for shallow circuits, however, as the depth of the circuit increases,
the performance is anticipated to degrade as bit-flip errors violate the constraint more frequently.
To overcome this problem, error correction, for example, as described in the Ref. \cite{Streif2}, will be important. Such efforts are the subject of future work.



\appendices
\renewcommand{\theequation}{\thesection.\arabic{equation}}
\thesection
\setcounter{equation}{0}
\section{\centerline{$D$-leg Ladder Driver Hamiltonian}}\label{sec:aplad}
Here, we show the driver Hamiltonian $\hat{\cal H}_d = \hat{\cal H}_{\rm hop}^{\rm lad}$ on the $D$-leg ladder lattice
and its ground state $\ket{\phi_0^{\rm lad}}$ used in the previous lad-FQAOA study \cite{FQAOA}.
The tight-binding model on the $D$-leg ladder lattice shown in \ref{fig:lattice} (b) is defined as:
\begin{align}
  \hat{\cal H}_{\rm hop}^{\rm lad}=&-t^{\parallel}\sum_{l=1}^N\left(\hat{c}_{l,d}^{\dagger}\hat{c}_{l+1,d}+\hat{c}_{l+1,d}^{\dagger}\hat{c}_{l,d}\right)\nonumber\\
  &-t^{\perp}\sum_{d=1}^{D-1}\left(\hat{c}_{l,d}^{\dagger}\hat{c}_{l,d+1}+\hat{c}_{l,d+1}^{\dagger}\hat{c}_{l,d}\right)\\
  =&\sum_{k=1}^{N}\sum_{m=1}^D\varepsilon^{\rm lad}_{k,m}\hat{\alpha}^{\dagger}_{k,m}\hat{\alpha}_{k,m},
  \label{eq:Hlad}
\end{align}
where $t^{\parallel}$ and $t^{\perp}$ are the longitudinal and transverse hopping integrals in the tight-binding model, respectively.
The lattice labels $l$ and $d$ are shown in Fig. \ref{fig:lattice} (b),
where the periodic boundary condition $\hat{c}_{N+1, d}=\hat{c}_{1,d}$ is imposed.

The energy dispersion $\varepsilon^{\rm lad}_{k,m}$ and the operator $\hat{\alpha}_{k,m}$ in Eq. (\ref{eq:Hlad}) have the following form \cite{FQAOA}:
\begin{equation}
  \varepsilon^{\rm lad}_{k,m}=-2t^{\parallel}\cos\left(\frac{2\pi k}{N}\right)-2t^{\perp}\cos\left(\frac{\pi m}{D+1}\right),
  \label{eq:eklad}
\end{equation}
and
\begin{equation}  
  \hat{\alpha}_{k,m}^\dagger = \sum_{l=1}^N\sum_{d=1}^D[{\bm \phi}_0^{\rm lad}]_{(k,m),(l,d)}\hat{c}_{l,d}^\dagger,\\
\end{equation}
respectively, with
\begin{equation}
\ [{\bm \phi}_0^{\rm lad}]_{(k,m),(l,d)}\\
=  \left \{\begin{alignedat}{2}
a_k\sin \left(\frac{2\pi kl}{N}\right)&\sin\left(\frac{\pi md}{D+1}\right),\\
&\text{if } 0< k < N/2,\\
a_k\cos \left(\frac{2\pi kl}{N}\right)&\sin\left(\frac{\pi md}{D+1}\right),\\
&\text{if } N/2\le k \le N,
\end{alignedat}
\right.\label{eq:phi0lad}
\end{equation}
where normalization factors $a_k = \sqrt{2/(D+1)N}$ for $k=N/2$ and $N$,
and $a_k = \sqrt{4/(D+1)N}$ for the other $k$'s.
The ground state of $\hat{\cal H}_d=\hat{\cal H}_{\rm hop}^{\rm lad}$,
the initial state of FQAOA, which is as follows:
\begin{equation}
  \ket{\phi_0^{\rm lad}} = \prod_{j=1}^{M}\hat{\alpha}_{(k,m)_j}^{\dagger}\ket{\text{vac}},
  \label{eq:phi0ladket}
\end{equation}
where the product for $k$ is taken such that $\sum_{j=1}^{M}\varepsilon^{\rm lad}_{(k,m)_j}$ is minimal.

The portfolio optimization problem discussed in the main text uses the parameters parameters $N=8$, $D=2$, and $K=4$.
In this case, the ground state is degenerate,
so we chose a symmetric occupied state as shown in following:
\begin{equation}
  \ket{\phi_0^{\rm lad}} = \hat{\alpha}_{N,2}^{\dagger}
  \hat{\alpha}_{1,1}^{\dagger}\hat{\alpha}_{N-1,1}^{\dagger}\hat{\alpha}_{N,1}^{\dagger}\ket{\text{vac}},
  \label{eq:phi0ladket}
\end{equation}
which is equivalent to the initial state used in the Ref. \cite{FQAOA}.

\section{Givens rotation gate used in present study}\label{sec:apgivens}
\setcounter{equation}{0}
As a supplementary note, we mention the implementation of a Givens rotation required for the initial state preparation.
In this study, to reduce the number of two-qubit gate operations,
the following circuit shown in Ref. \cite{IonQ} is adopted:
\begin{flalign}
\hspace{-10em}  
\Qcircuit @C=0.5em @R=.7em {
&\lstick{i}&\multigate{1}{{\cal G}(\theta)}&\qw&\\
&\lstick{i+1}&\ghost{{\cal G}(\theta)}&\qw&\push{\rule{3em}{0em}}}\nonumber
\end{flalign}
\begin{align}
  \begin{split}
    \Qcircuit @C=0.5em @R=0.7em {
&&&
  & \lstick{i}  &\gate{S}&\qw& \targ      & \gate{R_y(\theta)}           & \targ  &\qw&\gate{S^{\dagger}}   & \qw       \\  
&\push{\rule{0.9em}{0em}}&\lstick{\raisebox{1.8em}{=}}&\push{\rule{1.7em}{0em}}
  & \lstick{i+1}&\gate{S}&\gate{H}& \ctrl{-1} & \gate{R_y(\theta)} & \ctrl{-1} &\gate{H}&\gate{S^{\dagger}}& \qw &.
}
  \end{split}
\end{align}
As shown in the Eqs. (\ref{eq:phi0cyc}) and (\ref{eq:phi0lad}),
the the fermion orbitals on the cyclic and ladder lattices are written in real numbers,
so the $R_z(\varphi)$ gate in the Ref. \cite{FQAOA} is not needed.

\section*{Acknowledgment}

T.Y. thanks H. Kuramoto for valuable discussions.
K.F. is supported by MEXT Quantum Leap Flagship Program (MEXTQLEAP) Grants No. JPMXS0118067394 and No. JPMXS0120319794. This work is supported by JST COI-NEXT program Grant No. JPMJPF2014.


\end{document}